\documentclass[12pt,hyper, notoc]{JHEP3}

\pdfoutput=1
\usepackage{amsmath,amssymb,calc}
\usepackage{bbm}
\usepackage{cite}
\newcommand\be{\begin{equation}}
\newcommand\ee{\end{equation}}
\newcommand{\bea}{\begin{eqnarray}}
\newcommand{\eea}{\end{eqnarray}}

\newcommand{\nn}{\nonumber}

\def\beq{\begin{equation}}
\def\eeq{\end{equation}}
\def\id{\protect{{1 \kern-.28em {\rm l}}}}

\def\unit{\relax{\rm 1\kern-.26em I}}

\title{Non Abelian Tachyon Kinks}

\author{Vincenzo Cal\`o, Gianni Tallarita and Steven Thomas \\ \\ 
Queen Mary University of London \\
Center for Research in String Theory \\
Department of Physics, \\
Mile End Road, London, E1 4NS, UK. \\ 
Email: \email{V.Calo@qmul.ac.uk, G.Tallarita@qmul.ac.uk, S.Thomas@qmul.ac.uk}}

\abstract{Starting from the action of two coincident non-BPS D9-branes, we investigate kink configurations of the 
U(2) matrix tachyon field. We consider both $Str$ and $Tr$ prescriptions for the trace over gauge indices  of the non-BPS action. Non-abelian tachyon condensation in the theory with $Tr$ prescription, and the resulting fluctuations about the kink profile, are shown to give rise to a theory of two coincident BPS D8-branes. This is a natural non-abelian generalization of Sen's mechanism of tachyon condensation on a single non-BPS D$p$-brane yielding a single BPS brane of codimesion one.
By contrast, starting with the $Str$ gauge trace prescription of the coincident non-BPS D9-brane action, such a generalization of Sen's mechanism appears problematic.}

\preprint{QMUL-PH-09-10}
\keywords{Tachyon condensation, D-branes}

\begin{document}

\section{Introduction}
Tachyon condensation has long been an interesting aspect of D-brane physics (for a comprehensive review see \cite{Sen:2004nf}). Study of the dynamics of 
open string tachyons has
provided a fertile arena for studying various aspects of non-perturbative string theory. Such tachyons arise quite naturally in the open string spectrum 
when one considers non-BPS D-branes in type IIA or IIB string theories. A growing body of research 
has developed in open string field theory (for a review see \cite{Taylor:2003gn} or \cite{Schnabl:2005gv, Ellwood:2006ba} for more recent works) 
boundary string field theory, (BSFT) \cite{Witten:1992qy, Witten:1992cr, Shatashvili:1993kk, Kutasov:2000qp, N, Kraus:2000nj, Takayanagi:2000rz} and various effective 
actions around the tachyon vacuum \cite{Garousi:2000tr, Garousi:2002wq, Garousi:2003ai, Garousi:2004rd, Garousi:2007fn} to demonstrate Sen's  results \cite{Sen:1998ii, Sen:1998sm, Sen:1998tt, Sen:1998ex, Sen:1999xm} concerning the
 fate of the open string vacuum in the presence of tachyons. One particularly interesting aspect of tachyon dynamics 
that is captured by the various effective descriptions is the existence of 
solitonic configurations of the tachyon field \cite{Horava:1998jy}, 
including singular tachyon kink profiles \cite{Sen:2003tm,Kim:2003ma,Thomas:2005fw,Kluson:2005hd} which describe codimension one BPS branes as well as  
more exotic objects such as vortex solutions in brane-antibrane systems.

In \cite{Sen:2003tm}, the world-volume theory of the singular kink soliton solution (suitably regularized) where a single real tachyon 
field `condenses' on a single  non-BPS D-brane in a flat background was investigated using the effective Dirac-Born-Infeld (DBI) framework.
 Remarkably, it was shown that the effective theory of fluctuations about the tachyon kink profile, that depends only on a single spatial world-volume
coordinate, are precisely those of a codimension one BPS brane. Furthermore, it was also shown that in brane-antibrane systems, in which a single complex 
tachyon field is present, vortex solutions to the equations of motion exist, that naturally depend on two spatial worldvolume coordinates.
 Analysis of the fluctuations in this case show that they describe a codimension two  BPS D-brane. 
 Monopole solutions in certain truncations of tachyon models have also been found and initial investigations suggest that the corresponding 
 effective theory of fluctuations about this background correspond to codimension three BPS D-branes \cite{Hashimoto:2001rj}.

 In this paper we wish to investigate the process of tachyon condensation starting from the effective description of two coincident non-BPS D9-branes
 as proposed by Garousi in \cite{Garousi:2000tr}. This theory describes a non-abelian version of the DBI action in which the tachyon field transforms in the adjoint representation of the $U(2)$ gauge symmetry of the coincident non-BPS D9-brane 
 world volume action. In the original construction of this action and its generalization to coincident non-BPS D$p$-branes, a standard trace 
 prescription (which we denote as $Tr$) was taken over the gauge indices. Another prescription, motivated by string scattering calculations (at least to 
 low orders in $\alpha'$ \cite{Tseytlin:1997csa, Myers:1999ps}) is to take the symmetrized trace (which we denote by $Str$) over gauge indices. In both cases the expression being traced over is
 the same but the $Str$ prescription results in significantly more complicated terms in the action compared to $Tr$. 
 
 The effective theory of coincident non-BPS D9-branes is the simplest example of a multiple non-BPS brane action since there are no 
 matrix valued coordinate fields present perpendicular to the branes. We shall show that singular tachyon profiles exist
 which can be regularized in a way that preserves the $U(2)$ symmetry. We will see that studying the most general 
 fluctuations about this profile yields precisely the non-abelian DBI action of two coincident D8-branes. The only caveat is that our proof relies on 
 assuming the standard  $Tr$  as opposed to the $Str$ prescription for tracing over gauge indices in the DBI action of both the non-abelian non-BPS D9-brane action and the non-abelian D8-brane action. Whilst it is possible that tachyon condensation in the non-BPS action using $Str$ could lead to the $Str$ form of the action for two coincident D8-branes \cite{Tseytlin:1997csa, Myers:1999ps}, the exact mechanism for this to happen seems beyond a straightforward 
 extension of the method Sen used in the case of a single non-BPS brane \cite{Sen:2003tm}. In this sense the $Str$ prescription presents a 
 challenge for non-abelian tachyon condensation and deserves further investigation.
 
 As a simple check of the non-abelian tachyon condensation we also consider the case of non-abelian tachyon kinks where the $U(2)$ symmetry is 
 spontaneously broken to $U(1)~\otimes~U(1)$. The resulting effective theory of fluctuations is shown to lead to the sum of two DBI actions of separate 
 BPS D8-branes, as expected.
 
 The structure of the paper is as follows. In section 2 we review and motivate the non-abelian DBI action of coincident non-BPS D9-branes.
 In section 3 we study regularized kink profiles in the matrix valued tachyon field that preserve the $U(2)$ symmetry  and derive the effective world volume 
 theory of its fluctuations. In this section we also discuss the issues of $Tr$ vs $Str$ prescriptions and why the latter seems problematic as far as
 tachyon condensation is concerned. In section 4 we extend these results to kink profiles that spontaneously break $U(2)\rightarrow U(1)\otimes U(1)$.
 Finally in section 5 we end with some conclusions.

\section{Non-BPS $D9$-branes effective action}

In this section we shall introduce an effective action for the coincident non-BPS D9-brane pair. This system is unstable and it contains a tachyon in its spectrum, in particular, around the maximum of the tachyon potential, the theory contains a $U(2)$ gauge field and four tachyon states represented by a $2\times 2$ hermitian matrix-valued scalar field transforming in the adjoint representation of the gauge group.

To arrive at an effective action for this system, we first consider the effective action of a D$p$-anti-D$p$-brane pair proposed in \cite{Sen:2003tm}. In this case, the gauge group is $U(1) \times U(1)$ and so there are two massless gauge fields $A_{\mu}^{(1)}$ and $A_{\mu}^{(2)}$, a complex tachyon field $T$ and scalar fields $Y^I_{(1)}$, $Y^I_{(2)}$ corresponding to the transverse coordinate of individual branes. In particular, the action reads 
\begin{equation}
\label{dpdpb}
S= - \int d^{p+1}x \, V( T, Y^I_{(1)}-Y^I_{(2)} ) \left( \sqrt{-\textrm{det}G_{(1)}}+ \sqrt{-\textrm{det}G_{(2)}} \right) 
\end{equation}
where
\be
G_{(i) \mu \nu} = \eta_{\mu \nu}+ 2 \pi \alpha^{\prime} F^{(i)}_{\mu \nu} + \partial_{\mu} Y^I_{(i)} \partial_{\mu} Y^I_{(i)} + \pi \alpha^{\prime} (D_{\mu} T)^*(D_{\nu} T) + \pi \alpha^{\prime} (D_{\nu} T)^*(D_{\mu} T)\,.
\ee
This action has the nice property of admitting a vortex solution whose world volume action is given by the DBI action of a stable D$(p-2)$-brane \cite{Sen:2003tm}.

In \cite{Sen:1998ex, Dasgupta:1999qu} it has been proposed that the effective action of the D$p$-anti-D$p$ pair can be derived from the effective action of two non-BPS D$p$-branes by projecting it with $(-1)^{F_L}$ where $F_L$ is the spacetime left-handed fermion number. In particular, in the case of coincident $D9$-anti-$D9$-branes, the action (\ref{dpdpb}) can be derived from the following action \cite{Garousi:2007fn}:
\begin{equation}
\label{TrAction}
S=-Tr \int d^{10}x \, V(T) e^{-\phi} \, \sqrt[]{-\textrm{det} \, ( g_{\mu\nu}\mathbbm{1}_2+B_{\mu\nu}\mathbbm{1}_2+\pi\alpha'(D_{\mu}TD_{\nu}T+D_{\nu}TD_{\mu}T)+2\pi\alpha'F_{\mu\nu} ) }
\end{equation}
It is this effective action that we are going to use in order to construct the non-abelian kink solution. In eq.~(\ref{TrAction}), $g_{\mu\nu}$, $B_{\mu\nu}$ and $\phi $ are respectively the spacetime metric, the antisymmetric Kalb-Ramond tensor and dilaton fields whereas $\mathbbm{1}_2$ is the $2 \times 2$ unit matrix.
The covariant derivative is defined to be $D_{\mu}T=\partial_{\mu}T-i[A_{\mu},T]$ and the field strength takes the usual form $F_{\mu\nu}~=~\partial_{\mu}A_{\nu}-\partial_{\nu}A_{\mu}-i[A_{\mu},A_{\nu}]$.
The tachyon kinetic term has been written in a symmetric form to make the integrand a Hermitian matrix \cite{Garousi:2007fn}. $V(T)$ is the tachyon potential and although its exact form is still unknown, there are different proposals in the literature. For instance, the one which is consistent with S-matrix element calculation is given by \cite{Pesando:1999hm}
\begin{equation}
\label{SmatrixPotential}
V(T)=T_{9} \, (1+\pi\alpha'm^{2}T^{2}+\frac{1}{2}(\pi\alpha'm^{2}T^{2})^{2}+O(T^{6}))
\end{equation}
with $T_9$  the tension of the D9-brane and $m^2 = - \frac{1}{2 \alpha^{\prime} }$ the tachyon mass. 
The one given by boundary string field theory (BSFT) computations is \cite{Kraus:2000nj,Takayanagi:2000rz}
\be
\label{BSFTpotential}
V(T) = T_9 \, e^{- \pi \alpha' m^2 \, T^2}\, .
\ee
In particular, the potential (\ref{SmatrixPotential}) can be obtained from (\ref{BSFTpotential}) by expanding the latter around the tachyonic vacuum, $T=0$.
Henceforth, we shall not be interested in any specific form of the tachyon potential and, following \cite{Sen:2003tm}, we shall only assume that
\begin{itemize}
\item $V(T)$ is symmetric under $T \rightarrow -T$,
\item $V(T)$ has a maximum at $T=0$ and its minima are at $T=\pm \infty$ where it vanishes.
\end{itemize}

Before concluding this section, let us mention that in \cite{Garousi:2007fn} another form of the effective action for a coincident non-BPS $D9$-brane pair has been proposed. It is given in terms of the symmetrized trace\footnote{$Str(M_1 \ldots M_n) \equiv Tr \sum_{\sigma} \, M_1 \ldots M_n $ where  $\sum_\sigma $  is a sum over all permutations of matrices in $M_1 \ldots M_n$ divided by $n!$. } \cite{Tseytlin:1997csa, Myers:1999ps} 
\begin{equation}
\label{StrAction}
S=- Str \, \int d^{10}x \, V(T) e^{-\phi} \, \sqrt[]{-\textrm{det} \, ( g_{\mu\nu}\mathbbm{1}_2+B_{\mu\nu}\mathbbm{1}_2+ 2\pi\alpha' D_{\mu}TD_{\nu}T+2\pi\alpha'F_{\mu\nu}) }
\end{equation}
Various couplings in this action are consistent with the appropriate disk level S-matrix elements in string theory. In the above action the $Str$ prescription means specifically that one has to first symmetrize over all orderings of terms like $ F_{\mu \nu}, D_\mu T$ and also individual $T$ that appear in the potential $V(T)$.
The $Tr$ or $Str$ forms of the action are thus very different when one has carried out the individual symmetrizations mentioned above. 
As we discussed before, by projecting this action with $(-1)^{F_L}$ one can obtain the effective action of a $D_9$-anti-$D_9$-brane pair. 
However, for this action there are no known solutions corresponding to a vortex whose world volume is given by the DBI action of a stable $D7$-brane.

\section{Non Abelian Kink}

To simplify our calculations we set $B_{\mu\nu}=0$, $g_{\mu\nu}=\eta_{\mu \nu}=(-1,1,\ldots, 1)$ and take a constant dilaton $\phi$ consistent with the flat background. We also set the gauge fields to zero. The latter will be reintroduced when we consider fluctuations around the kink solution.

\subsection{Energy-momentum tensor and equations of motion}

In this section we shall compute the energy-momentum tensor and the equations of motion associated with the actions (\ref{TrAction}) and (\ref{StrAction}). 
In particular the energy-momentum tensor associated with the action (\ref{TrAction}) is given by
\begin{equation}
T_{\mu\nu}=-Tr \, V(T) \, \sqrt{-detG} \, G^{-1}_{\mu\nu}
\end{equation}
where we defined
\begin{equation}
G_{\mu\nu} \equiv  \eta_{\mu\nu}+B_{\mu\nu} +\pi\alpha' (D_{\mu}TD_{\nu}T+D_{\nu}TD_{\mu}T)+2\pi\alpha'F_{\mu\nu}  \,.
\end{equation}
A similar expression holds for the symmetrized trace form of the action but with $Tr$ replaced by $Str$.

Following Sen \cite{Sen:2003tm}, we show that the kink solution consistent with the energy-momentum conservation and the e.o.m is given by
\begin{equation}
\label{Tansatz}
T(x)=f(a\frac{x}{\sqrt{\alpha'}}) \mathbbm{1}_2 =f(a\frac{x}{\sqrt{\alpha'}} \, \mathbbm{1}_2)  
\end{equation}
with gauge fields set to zero,  $x \equiv x^9$ a direction longitudinal to the system and $a$ an arbitrary dimensionless constant that we should take to infinity at the end. The function $f(u)$ can be any real function with the property that $f(u \rightarrow \pm \infty) \rightarrow \pm \infty $ and $f'(u)>0$, $\forall u$. 
As a matter of fact, eq. (\ref{Tansatz}) is a way of regularizing the tachyon singular solution which comes from the energy-momentum conservation condition $\partial_x T_{xx}=0$: the latter implies that 
\be
T_{xx}= -Tr \, \frac{V(T)} {\sqrt{1+2 \pi \alpha' \partial_x T\partial_x T}}
\ee
must be independent of $x$. Therefore, since for $x~\rightarrow~\infty$ we have that $T_{xx} \rightarrow 0$ then\footnote{Recall that for a kink solution $\lim_{x \rightarrow \infty} T \rightarrow \infty$ and we assumed that the tachyon potential is zero at infinity.} $T_{xx}~ =~0$, $\forall \, x$.
We conclude that $T$ is singular, namely
\begin{equation}
\label{singularity}
T=\pm \infty \quad \textrm{and/or} \quad \partial_{x} T =\pm \infty \quad \forall \,x \, 
\end{equation}
and this singularity is regularized by taking the constant $a$ in (\ref{Tansatz}) to infinity.
However, one can also show that this kink solution has finite energy density regardless of the way of regularizing the singularity.

Let's compute now the equation of motion for the tachyon (keeping the gauge fields non-zero), in particular, varying eq. (\ref{Tansatz}) w.r.t. $T$ we obtain:
\begin{equation}
\label{TrEOM}
\pi\alpha'D_{\rho}\left(V(T)\sqrt{-detG} \, (G^{-1})^{\mu\nu}(D_{\nu}T\delta_{\mu}^{\rho}+D_{\mu}T\delta_{\nu}^{\rho})\right)-\frac{\partial V(T)}{\partial T}\sqrt{-detG}=0
\end{equation}
where we use the properties of the trace to permute all the various sources of $\delta T$ factors that arise in the variation of the action. When one uses the symmetrized trace form of the action (\ref{StrAction}) the equations of motion for $T$ are:
\begin{equation}
\label{StrEOM}
\Sigma_{\sigma} \left[\pi\alpha'D_{\rho}\left(V(T)\sqrt{-detG} \,  (G^{-1})^{\mu\nu} (D_{\nu}T\delta_{\mu}^{\rho}+D_{\mu}T\delta_{\nu}^{\rho})\right)-\frac{\partial V(T)}{\partial T} \sqrt{-detG} \right]=0
\end{equation}
where $\sum_{\sigma} $  accounts for all symmetrical permutations of the matrices inside the squared brackets in the previous expression.

We now verify that the kink solution eq. (\ref{Tansatz}) satisfy the equation of motions (\ref{TrEOM}) in the $a \rightarrow \infty$ limit. In this case:
\begin{equation}
G_{\mu\nu}= \eta_{\mu\nu}+2\pi\alpha' \partial_{\mu}T\partial_{\nu}T  = \left(\begin{array}{cccc}
            -1&0&\ldots&0\\ 0&1&\ldots& 0 \\
             \vdots &\ddots & \ddots& \vdots \\ 0& \ldots & 0 & (1+2a^2 \pi (f^{'})^{2})
             \end{array}\right) \otimes \mathbbm{1}_2
             \end{equation}
where $'$ denotes differentiation w.r.t. the dimensionless argument of $f$. 
It follows that
\begin{equation}
\label{detG}
- det G = 1+2a^2 \pi(f^{'})^{2}  \approx 2a^2 \pi(f^{'})^{2} 
\end{equation}
and 
\begin{equation}
\label{InvG}
(G^{-1})^{\mu\nu}= \left[\eta^{\mu\nu}+\left(\frac{1}{1+2 a^2 \pi (f^{'})^{2}}-1\right) \delta_{x}^{\mu}\delta_{x}^{\nu} \right] \otimes \mathbbm{1}_2 \,.
\end{equation}
%In the previous expressions and henceforth, the prime $'$ denotes differentiation w.r.t. $x$.
Substituting eqs.~(\ref{Tansatz}), (\ref{detG}) and (\ref{InvG})  into eq. (\ref{TrEOM}) one obtains
\begin{eqnarray}
  &&2\pi  \alpha'\partial_{x}\left( V(T) \sqrt{-detG} \, (G^{-1})^{xx}  \partial_x T \right)  - \frac{\partial V(T)}{\partial T} \sqrt{-detG} \nonumber \\
  &=&2 \pi\sqrt{\alpha'}\partial_{x}\left( V(T) \frac{1}{\sqrt{1+2 a^2 \pi (f^{'})^{2}}} a f^{'} \right) - \frac{\partial V(T)}{\partial T}  \sqrt{1+2 a^2 \pi (f^{'})^{2}}   \nn \\ 
  &\approx& \sqrt{2 \pi \alpha'} \partial_x V(T) -  \sqrt{2 \pi}  a f^{'} \frac{\partial V(T)}{\partial T}  =0
\end{eqnarray}
where in the last step we have taken the large $a$ limit.
Notice that since the solution (\ref{Tansatz}) is such that both $T$ and $D_x T$ commute (indeed they are both proportional to the identity in group space), then it is equally a solution of the equations of motion derived from the $Str$ procedure eq.~(\ref{StrAction}) in the background in which the gauge fields are set to zero.

\subsection{Study of the fluctuations}

We proceed to study the fluctuations around the solution (\ref{Tansatz}) which preserve the $U(2)$ symmetry. 
These fluctuations correspond just to shifts in the argument of the function $f(a\frac{x}{\sqrt{\alpha'}})$. The analysis is similar to \cite{Sen:2003tm}, however, we now have two copies of the usual abelian tachyon profile filling out the diagonal elements of the matrix tachyon field, thus representing the two coincident D8-branes.

\subsubsection{Warmup: $ T =  f(\frac{a}{\sqrt{\alpha'}} (x -t(\xi))) \mathbbm{1}_2$}
As a warmup calculation we consider a fluctuation of the type
\begin{equation}
\label{flucIdentity}
 T= f(\frac{a}{\sqrt{\alpha'}}(x_{1}-t(\xi))) \mathbbm{1}_2\,,
 \end{equation}
where $\xi^{\alpha}$ denotes all the coordinates tangential to the kink world-volume and $t(\xi)$ the field associated with the translational zero mode of the kink. 
Taking the group trace, $Tr$ or $Str$, in the action (\ref{TrAction}) or (\ref{StrAction}) in the case where the tachyon profile 
and its derivatives are proportional to the identity as in eq.~(\ref{flucIdentity}), will thus give us two identical D8-brane actions\footnote[1]{Note that in the determinant under the square root the symmetric $D_{\mu}TD_{\nu}T$ term is
 automatically diagonalized in the gauge indices.}. Indeed, for the fluctuation (\ref{flucIdentity}),
 \begin{equation}
-detG=1+ 2 \pi a^2 (f')^{2}\, (1+\eta^{\alpha\beta}\partial_{\alpha}t\partial_{\beta}t)
\end{equation}
and we obtain
\bea
S&=&-Tr\int d^{9}\xi \, dx \, V(f)\, \sqrt{2 \pi} a f' \sqrt{1+\eta^{\alpha\beta}\partial_{\alpha}t\partial_{\beta}t} \cr
&=&-2 \sqrt{2 \pi} a \int d^{9}\xi \, dx \, V(f)  f'\sqrt{1+\eta^{\alpha\beta}\partial_{\alpha}t\partial_{\beta}t}
\eea
and by a substitution $y= f(\frac{a}{\sqrt{\alpha'}}(x-t(\xi)))$ one finds
\begin{equation}
S=-2 \sqrt{2 \pi \alpha'}\int^{\infty}_{-\infty}dy V(y) \, \int d^{9}\xi \sqrt{1+\eta^{\alpha\beta}\partial_{\alpha}t\partial_{\beta}t}
\end{equation}
which upon the identification $\textit{T}_{8}=\sqrt{2 \pi \alpha'}\int^{\infty}_{-\infty}dy V(y)$ we recognize as the action describing two identical D8-branes (with no separation) with a single translational fluctuation mode $t(\xi)$ turned on.

\subsubsection{$ T = f(\frac{a}{\sqrt{\alpha'}}(x \mathbbm{1}_2 -t^a(\xi) \sigma_a)) $}

Of course it is well known that the full DBI action for coincident BPS D8-branes
should involve a nonabelian theory in which the single coordinate perpendicular to the D8-brane worldvolume is a $U(2)$ matrix-valued field and the resulting action has local $U(2)$ gauge invariance.
Thus we would like to show how such an action appears by looking at the most general fluctuations around 
our original kink solution $T = f(\frac{a}{\sqrt{\alpha'}}x) \mathbbm{1}_2$.
To this end, let us keep the fluctuations in the gauge field zero for the time being and consider fluctuations of the tachyon profile of the form:
\begin{equation}\label{flucGeneral}
T= f(\frac{a}{\sqrt{\alpha'}}(x \mathbbm{1}_2-t^{a}(\xi)\sigma_{a}))
\end{equation}
where $\sigma^a = (\sigma^0=\mathbbm{1}_2, \sigma^i)$, $\sigma^i$ being the Pauli matrices and we should regard $f$ as a matrix-valued application expressed as an infinite power series of its argument.
The above ansatz for the fluctuations is a natural non-abelian generalization of the one that Sen used to describe fluctuations of regularized tachyon kink in the abelian case \cite{Sen:2003tm}.  

If in the first instance, we make use of the quadratic approximation for the determinant:
\begin{equation}
detG_{\mu\nu}=\mathbbm{1}_2 +2 \pi \alpha^{'} \partial_{\mu}T\partial^{\mu}T +{\cal O}(\alpha^{'2})
\end{equation}
the action in the large $a$ limit becomes
\begin{equation}
\label{FlucAction1}
S=-Tr \int d^{10}x V(f) \sqrt{2 \pi} a\sqrt{f'^{2}} \sqrt{\mathbbm{1}_2+  \partial_{\alpha}t\partial^{\alpha}t}
\end{equation}
where $t$ is the $U(2)$ matrix $t^{a}\sigma_{a}$. 

In obtaining the above we have implicitly assumed that $\partial_\alpha f = -\frac{a}{\sqrt{\alpha'}} f' \partial_\alpha t$ while 
$\partial_{x}~f~=~\frac{a}{\sqrt{\alpha'}}~f'$ is identically the case since the dependence on $x$ is via the unit matrix $\mathbbm{1}_2$ in $f$.
In fact, there is a subtlety associated with the former relation: since $\partial_\alpha t$ and $t$ do not commute in general,
 there is an ordering issue that means that for general functions $f$, differentiating w.r.t. 
 $\xi^\alpha$ one cannot simply use the chain rule and express the result as $-\frac{a}{\sqrt{\alpha'}}f'\partial_\alpha t$. 
 There will be various symmetric ordering of $\partial_\alpha t$ and $t$ that spoil this. 

However there is at least one example, namely when $f(u)$ is linear in its argument (with positive coefficient so that 
$f'>0$ everywhere as required) where the chain rule will hold and no ordering problems occur when differentiating.

The linear form of $f$ has another interesting feature. If we had started with the $Str$ form of the action, then
as discussed above this implies symmetrization w.r.t. $ F_{\mu \nu}, D_\mu T$ and $T$. For linear $f$ we see that 
it follows that this $Tr$ procedure immediately implies a similar $Str$ procedure where we replace $T$ with $t$.
This is exactly what we would expect if we require that the $Str$ procedure is the one that correctly describes coincident
D8-branes with $t$ the single transverse coordinate to the world volume.

Finally it is interesting to observe that as pointed out in \cite{Sen:2003tm}, the linear tachyon profile seems to play and important role in the BSFT 
description of tachyon vortex solutions discussed in \cite{Kraus:2000nj, Takayanagi:2000rz}.

For all these reasons the linear form of $f$ seems to be singled out as being special. For now we will leave $f$ in its generic
form but bear in mind these issues.

The action (\ref{FlucAction1}) looks  of the right form, i.e., it is a non-abelian DBI action (though with the gauge field fluctuations yet to be included). 
However, one faces taking the square root of the function $f'^{2}$ which is matrix valued and is thus non trivial. 
One has to diagonalize the matrix $f$ first in order to take its square root and obtain a closed form expression. 
The terms inside the second square root part of the action are proportional to the identity and so we can diagonalize them by a $U(2)$ 
transformation directly:
\begin{eqnarray}
S&=&- \sqrt{2\pi} a\, Tr \int d^{10}x \,V(f) \, \sqrt{f'^{2}} \sqrt{\mathbbm{1}_2+\partial_{\alpha}t\partial^{\alpha}t}\cr
&=&- \sqrt{2\pi} a\, Tr \int d^{10}x \,U^{\dagger}\, V(f) U\, U^{\dagger} \sqrt{f'^{2}} U \, \sqrt{\mathbbm{1}_2+\partial_{\alpha}t\partial^{\alpha}t}\cr
&=&- \sqrt{2\pi} a\, Tr \int d^{10}x \, V( U^{\dagger} f U ) \, \sqrt{U^{\dagger}f'^{2}U} \, \sqrt{\mathbbm{1}_2+\partial_{\alpha}t\partial^{\alpha}t} \,.
\label{FlucAction2}
\end{eqnarray}
Now,
\begin{eqnarray}
U^{\dagger}f(\frac{a}{\sqrt{\alpha'}}(x \mathbbm{1}_2+t^{a}(\xi)\sigma_{a}))U&=&f(U^{\dagger}\frac{a}{\sqrt{\alpha'}}(x \mathbbm{1}_2+t^{a}\sigma_{a})U)=
f\left( \frac{a}{\sqrt{\alpha'}} \left( (x +t^{0})\mathbbm{1}_2+U^{\dagger}t^{i}\sigma_{i}U \right) \right)\cr
&=&f\left(\frac{a}{\sqrt{\alpha'}} \left( (x+t^{0})\mathbbm{1}_2+\sqrt{t^{a}t_{a}}\sigma_{3} \right) \right) \,.
\end{eqnarray}
This diagonalization then describes a matrix of the form:
\begin{eqnarray}
U^{\dagger}f(\frac{a}{\sqrt{\alpha'}}(x \mathbbm{1}_2+t^{a}(\xi)\sigma_{a}))U&=& \left(\begin{array}{cc}
f \left( \frac{a}{\sqrt{\alpha'}} ( x+t^{0}+\sqrt{t^{a}t_{a}}) \right) &0\\
0&f\left( \frac{a}{\sqrt{\alpha'}} (x+t^{0}-\sqrt{t^{a}t_{a}} )  \right)\end{array} \right) \nn \\
&&\nn\\
&\equiv& {\cal D} (f_{1},f_{2})
\end{eqnarray}
where 
\begin{eqnarray}
f_{1}&=&f \left( \frac{a}{\sqrt{\alpha'}} ( x+t^{0}+\sqrt{t^{a}t_{a}}) \right)\,,\nn \\ f_{2}&=&f \left( \frac{a}{\sqrt{\alpha'}} ( x+t^{0}-\sqrt{t^{a}t_{a}}) \right)\,. \nonumber
\end{eqnarray} 
We also note that the matrix used to diagonalize $f$ only depends on the variables $t^{i}(\xi)$ which means that $U^{\dagger}f'U=(U^{\dagger}fU)'$ and so the action (\ref{FlucAction2}) becomes
\begin{eqnarray}
S&=&- \sqrt{2\pi} a \, Tr \int d^{10}x  \, {\cal D}(V(f_{1}), V(f_{2})) \, {\cal D}({f'_1},{f'_2}) \sqrt{\mathbbm{1}_2+\partial_{\alpha}t\partial^{\alpha}t} \cr
&=&-\sqrt{ 2\pi} a \, Tr \int d^{10}x \, {\cal D}(V(f_{1}){f'_1}, V(f_{2}){f'_2}) \, \sqrt{\mathbbm{1}_2+\partial_{\alpha}t\partial^{\alpha}t} \,.
\end{eqnarray}
Substituting for the variables $y=f_{1}$ and $z=f_{2}$ we obtain the generalization of Sen's procedure for the non-abelian case:
\begin{eqnarray}\label{d8action}
S&=&- \sqrt{2 \pi \alpha'}Tr \int d^{9}x \, {\cal D} \left(\int^{\infty}_{-\infty} dy V(y)\, ,\int^{\infty}_{-\infty} dz V(z)\right)\sqrt{\mathbbm{1}_2+\partial_{\alpha}t\partial^{\alpha}t}\cr
&=&- T_{8} \, Tr\int d^{9}x \sqrt{\mathbbm{1}_2+\partial_{\alpha}t\partial^{\alpha}t}
\end{eqnarray}
which we recognize as the non-abelian DBI action for the coincident D8-branes (with gauge fields set to zero) upon identifying the 
tension $T_{8}=\sqrt{2 \pi \alpha'}\int^{\infty}_{-\infty} dy V(y)$.
 In order to be sure that in the $a \rightarrow \infty$ limit one really is in the vacuum of the theory we must look at 
 the potential for the matrix form of $T$:  the requirement that $V(f(\pm\infty))=0$ is enough to ensure that.

Now one might also try and arrive at the $Str$ form of the above action, by starting with the $Str $ form of the tachyon action for 
non-BPS D9-branes (\ref{StrAction}).  The terms inside the square root part of the action are diagonal in $U(2)$ space and so one can imagine expanding 
out the square root factor in a power series and them symmetrizing over terms involving $\partial_\alpha T $ and $T$ in $V(T)$.
The problem one encounters then is that integrating over $dx$ by making the change of variables as above does not look
feasible due to the non-commutation between $f$ and $\partial_\alpha t$ terms. That is, even using the cyclic properties of $Tr$, terms obtained 
through $Str$ cannot be factorized into terms involving just powers of  $f$ times those involving $\partial_\alpha t$. 
Therefore, it seems that a straightforward generalization of Sen's procedure to show that non-abelian tachyon condensation 
via kink solitons in coincident non-BPS brane theories gives rise to coincident D$p$-branes is only possible in the $Tr$ 
prescription rather than $Str$. It is interesting to see here a parallel to the problem of $Str$ vs $Tr$ prescriptions in trying to realize 
 vortex (as opposed to kink) solutions in brane-antibrane systems obtained from coincident non-BPS D9-branes \cite{Garousi:2007fn}.

Working within the $Tr$ prescription, let us now proceed to include the gauge field fluctuations and to go beyond the quadratic 
approximation of the determinant used before, to include all higher order terms. We take the following ansatz for the gauge fields \cite{Sen:2003tm}:
\begin{eqnarray}
\label{GaugeFieldAnsatz}
A_{x}(x,\xi)=0 \,,&&A_{\alpha}(x,\xi)=a(\xi)_{\alpha}^{a}\sigma_{a}\,,
\end{eqnarray}
%where $\alpha$ labels the indices tangent to the world volume of the kink-like solution. 
%and we rewrite the tachyon kinetic term in (\ref{TrAction}) as
%\begin{equation}
%D_{\mu}TD_{\nu}T+D_{\nu}TD_{\mu}T=2 D_{\mu}T^{a}D_{\nu}T^{b}d_{abc}\sigma_{c}
%\end{equation}
%where $d_{abc}$ denotes the totally symmetric structure constants of the $U(2)$ algebra under anticommutation.
 
Now let us pause briefly to comment on the action of the covariant derivative $D_\alpha$ on the function $f$ 
 appearing in the ansatz eq.~(\ref{flucGeneral}) for the tachyon kink. Just as we mentioned earlier when discussing 
 the action of $\partial_\alpha$ on $f$, the commutator terms $[A_\alpha,f]$ cannot, in general, easily be expressed in 
 terms of $f'$ and $[A_\alpha, t]$ which is what we would have hoped if we are to promote the action eq.~(\ref{d8action})
  to one that is locally $U(2)$ invariant. There are again ordering issues arising form the non-commutativity of $[A_\alpha,t]$ 
  and $t$. Taking  $f(u)$ linear in its argument avoids this as before. For now let us just keep $f$ in our expressions but have 
  in mind that it is likely to be constrained to be linear if we assume that $D_{\alpha}T=-\frac{a}{\sqrt{\alpha'}}f'D_{\alpha}t$.

We can proceed with calculating the determinant of the matrix in the action using the ansatz (\ref{flucGeneral}) for the tachyon 
field and (\ref{GaugeFieldAnsatz}) for the gauge fields. We obtain 
\begin{eqnarray}
G_{xx}&=&(1+2\pi a^{2}f'^{2})\\
G_{\alpha x}&=&-2\pi a^{2}f'^{2}D_{\alpha}t\\
G_{\alpha\beta}&=&\pi a^{2}f'^{2}(D_{\alpha}tD_{\beta}t+D_{\beta}tD_{\alpha}t)+a_{\alpha\beta}
\end{eqnarray}
where $a_{\alpha\beta}=\eta_{\alpha\beta}+2\pi\alpha'F_{\alpha\beta}$. Now we can make use of Sen's trick \cite{Sen:2003tm} of adding rows and columns of the same matrix to simplify the computation of the determinant. In particular, we have
\begin{eqnarray}
\hat{G}_{\mu\beta}&=&G_{\mu\beta}I_{2}+\frac{1}{2}G_{\mu x}D_{\beta}t+\frac{1}{2}D_{\beta}tG_{\mu x}\\
\hat{G}_{\mu x}&=&G_{\mu x}
\end{eqnarray}
and finally:
\begin{eqnarray}
\tilde{G}_{\alpha\nu}&=&\hat{G}_{\alpha\nu}I_{2}+\hat{G}_{x\nu}D_{\alpha}t\\
\tilde{G}_{x\nu}&=&\hat{G}_{x\nu}
\end{eqnarray}
from which we obtain 
\begin{equation}
\tilde{G}_{xx}=(1+2\pi a^{2}f'^{2})\mathbbm{1}_2 , \quad \tilde{G}_{x\alpha}=\tilde{G}_{\alpha x}=D_{\alpha}{t(\xi)}^{a}\sigma_{a} , \quad \tilde{G}_{\alpha\beta}=\tilde{a}_{\alpha\beta}
\end{equation}
where 
\begin{equation}\label{atilde}
\tilde{a}_{\alpha\beta}=a_{\alpha\beta}+D_{\alpha}t^a(\xi) D_{\beta}t^b(\xi)\sigma_{a}\sigma_{b} \,.
\end{equation}
This means that overall
\begin{equation}
\textrm{det} (\tilde{G}_{\mu\nu})= \textrm{det} (G_{\mu\nu})= 2 \pi a^{2}f'^{2}det(\tilde{a}_{\alpha\beta})+O(\frac{1}{a^{2}}) \,.
\end{equation}
The last equation is precisely the generalization of the result Sen obtained to the case of local $U(2)$ gauge covariant quantities.
Note that in the above manipulations we have taken $f'$ to commute through expressions involving $U(2)$ matrices. 
For general $f$ this would not be the case but for linear $f$, $f'$ is simply proportional to the $2\times 2 $ identity matrix 
as noted earlier, so this is justified. 

 We can now substitute this result into the action to obtain
\begin{eqnarray}
S&=&-\sqrt{2 \pi} a \, Tr \int d^{10}x \, {\cal D} (V(f_{1})f'_{1},V(f_{2})f'_{2}) \sqrt{- det(\tilde{a}_{\alpha\beta})}
\end{eqnarray}
which is the full non-abelian DBI action for two coincident D8-branes (using the $Tr$ prescription) once the usual parameter substitutions 
are performed and the resulting integral over $x$ identified with the D8-brane tension $T_8$:
\begin{eqnarray}\label{2D8}
S&=&-T_8 \, Tr  \int d^9x \sqrt{-det(\tilde{a}_{\alpha\beta})} \,.
\end{eqnarray}

Now one should also show, as a further check, that the solutions of the equations of motion arising from the
 action (\ref{2D8}) coincide with the solutions as derived from the original coincident non-BPS D9-brane action ({\ref{TrAction}),
  upon using the non-abelian tachyon profile given in eq.~(\ref{flucGeneral}). This check was done explicitly by Sen in \cite{Sen:2003tm} 
  in the case of tachyon condensation on a single non-BPS D$p$-brane. The calculation in our case would follow quite closely that of Sen, 
  just extended to the non-abelian case relevant to two coincident D-branes. The main points of the proof use the property that  
  $D_\alpha f = -\frac{a}{\sqrt{\alpha'}} f' D_\alpha t$ used earlier and the approximate relation 
  $ det(G_{\mu\nu}) = 2 \pi a^{2}f'^{2}det(\tilde{a}_{\alpha\beta})+O(\frac{1}{a^{2}}) $.
  Details will be presented elsewhere \cite{CTT:2009}.

\section{Breaking $U(2)$ to $U(1) \otimes U(1)$}

As further check on our generalized Sen ansatz eq.~(\ref{flucGeneral}), we can consider
modifying the argument of $f$ so that the corresponding kink solution breaks $U(2)$ symmetry and thus should describe a pair of separated D8-branes after condensation. This amounts to  
 allowing  a vacuum expectation value to one of the $U(2)$ adjoint fields $t^i$. In particular, we set $t(\xi) \rightarrow t(\xi)+c \sigma_{3}$, where $c$ denotes a constant v.e.v. related to the separation of the two D8-branes along their single transverse direction. In this case we expect to break the $U(2)$ invariance of the theory down to $U(1)\otimes U(1)$. The resulting action of fluctuations about this vacuum configuration should split into two abelian DBI actions, i.e., two distinct determinant terms each carrying a single $U(1)$ gauge field and perpendicular scalar fluctuation field, that describe the separate D8-branes.

We start by introducing the v.e.v. $c$ and obtain a modification of eq.~(\ref{atilde}) due to this shift: in particular
\begin{eqnarray}
\tilde{G}_{\alpha\beta}=\tilde{a}_{\alpha \beta}&=&a_{\alpha\beta}+\partial_{\alpha}t\partial_{\beta}t-i\partial_{\alpha}t[A_{\beta},t]-i[A_{\alpha},t]\partial_{\beta}t-[A_{\alpha},t][A_{\beta},t] \cr &&
-ic \, \partial_{\alpha}t \,[A_{\beta},\sigma_{3}]-ic[A_{\alpha},\sigma_{3}]\partial_{\beta}t- c[A_{\alpha},t][A_{\beta},\sigma_{3}] -c[A_{\alpha},\sigma_{3}][A_{\beta},t] \cr
&&-c^{2}[A_{\alpha},\sigma_{3}][A_{\beta},\sigma_{3}]
\end{eqnarray}
where the covariant derivatives appearing in eq.~(\ref{atilde}) have been expanded out explicitly. To proceed we make use of a different parametrization of $t$ that makes explicit the Goldstone modes associated with $U(2)$ symmetry breaking: we set
\begin{equation}
t^{a}\sigma_{a}=U^{\dagger}(\tilde{t}^{0}\mathbbm{1}_2+\tilde{t}^{3}\sigma_{3})U
\end{equation}
where $U=\exp^{\frac{i}{c}(\tilde{t}^{1}\sigma_{1}+\tilde{t}^{2}\sigma_{2})}$ and we pick a preferential gauge in which
\begin{eqnarray}
(t^{a}\sigma_{a})' &=& Ut^{a}\sigma_{a}U^{\dagger}=\tilde{t}^{0}\mathbbm{1}_2+\tilde{t}^{3}\sigma_{3} \\
(A_{\alpha}^{a}\sigma_{a})'&=&U(A_{\alpha}^{a}\sigma_{a})U^{\dagger}-(\partial_{\alpha}U)U^{\dagger} \, .
\end{eqnarray}
In this gauge, the fluctuations $t$ are diagonal and\footnote{We drop the prime sign from the gauged form of $A_{\alpha}'$ and the tilde on  $\tilde{t}^0, \tilde{t}^3$.}
\begin{eqnarray}
\partial_{\alpha}t\partial_{\beta}t &=& (\partial_{\alpha}t^{0}\partial_{\beta}t^{0}+\partial_{\alpha}t^{3}\partial_{\beta}t^{3}) \mathbbm{1}_2+( \partial_{\alpha}t^{0}\partial_{\beta}t^{3}+\partial_{\alpha} t^{3}\partial_{\beta}t^{0}) \sigma_{3} \cr
\partial_{\alpha}t\, [A_{\beta},t] &=& 2 i t^3 \partial_\alpha t^0 \left( A^2_\beta \sigma_1 - A^1_\beta \sigma_2\right)- 2 t^3 \partial_\alpha t^3 \left( A^2_\beta \sigma_2 + A^1_\beta \sigma_1\right) \cr
[A_{\alpha},t][A_{\beta},t] &=& 4 (t^3)^2 \left( -A_{\alpha}^1 A^1_\beta - A_{\alpha}^2 A^2_\beta +i \left(A_{\alpha}^2 A^1_\beta -A_{\alpha}^1 A^2_\beta  \right)\sigma_3 \right)
\end{eqnarray}
with similar expressions holding with various $t^3$ are replaced by the v.e.v. $c$. Now we redefine the gauge fields so as to absorb the v.e.v. $c$ by setting $A^i_{\alpha} = \frac{1}{2 c} \tilde{A}^{i}_\alpha$ for $i=1,2$. 
Substituting these expressions and taking the large $c$ limit one obtains to leading order
\begin{eqnarray}
\tilde{G}_{\alpha\beta} &=& \eta_{\alpha\beta}+ F_{\alpha\beta}^{0} \mathbbm{1}_2+F_{\alpha\beta}^{3}\sigma_{3}+(\partial_{\alpha}t^{0}\partial_{\beta}t^{0}+\partial_{\alpha}t^{3}\partial_{\beta}t^{3}) \mathbbm{1}_2+( \partial_{\alpha}t^{0}\partial_{\beta}t^{3}+\partial_{\alpha} t^{3}\partial_{\beta}t^{0}) \sigma_{3}  \cr &&
\left(\partial_\alpha t^0 (A^2_\beta \sigma_1 -A^1_\beta \sigma_2) + (\alpha \leftrightarrow \beta) \right) + i \left(\partial_\alpha t^3 (A^1_\beta \sigma_1 + A^2_\beta \sigma_2) - (\alpha \leftrightarrow \beta) \right) \cr &&
 +(A_{\alpha}^1 A^1_\beta +A_{\alpha}^2 A^2_\beta)\mathbbm{1}_2 -i \left(A_{\alpha}^2 A^1_\beta -A_{\alpha}^1 A^2_\beta  \right)\sigma_3
\end{eqnarray}
The fields $A^{i}_{\alpha}, i=1,2 $ are non-propagating to lowest order in a $1/c $ expansion and a consistent solution of their equations of motion is to set  $A^{1}_{\alpha}=A^{2}_{\alpha}=0$. The limit of large $c$ corresponds to considering the two coincident D8-branes as being separated by a distance that is large compared to the string length $\sqrt{\alpha'}$.

 We use this and redefine the field strengths and scalar fields associated to each brane as
  $F^{1}_{\alpha\beta}=F_{\alpha\beta}^{0}+ F_{\alpha\beta}^{3}$, $F^{2}_{\alpha\beta}=F_{\alpha\beta}^{0}-F_{\alpha\beta}^{3}$ and 
  $\phi^{1}=t^{0}+t^{3}$, $\phi^{2}=t^{0}-t^{3}$. Then, in group space the matrix $\tilde{G}_{\alpha\beta}$ reduces to 
\begin{equation*}
\tilde{G}_{\alpha\beta} = \left(\begin{array}{cc}
									\eta_{\alpha\beta}+F^{1}_{\alpha\beta}+\partial_{\alpha}\phi^{1}\partial_{\beta} \phi^{1}&0\\
0&\eta_{\alpha\beta}+F^{2}_{\alpha\beta}+\partial_{\alpha}\phi^{2}\partial_{\beta}\phi^{2}
\end{array}\right)
\end{equation*}
hence,
\begin{equation*}
\sqrt{-\textrm{det}(\tilde{G}_{\alpha\beta})}=\left(\begin{array}{cc}
									\sqrt{-det(\eta_{\alpha\beta}+F^{1}_{\alpha\beta}+\partial_{\alpha}\phi^{1}\partial_{\beta}\phi^{1})}&0\\
0&\sqrt{-det(\eta_{\alpha\beta}+F^{2}_{\alpha\beta}+\partial_{\alpha}\phi^{2}\partial_{\beta}\phi^{2})}
\end{array}\right)
\end{equation*}
and finally defining $\tilde{G}^{1}_{\alpha\beta}=\eta_{\alpha\beta}+F^{1}_{\alpha\beta}+\partial_{\alpha}\phi^{1}\partial_{\beta}\phi^{1}$ and $\tilde{G}^{2}_{\alpha\beta}=\eta_{\alpha\beta}+F^{2}_{\alpha\beta}+\partial_{\alpha}\phi^{2}\partial_{\beta}\phi^{2}$ we find that the action becomes
\begin{equation}
S=-\sqrt{2 \pi}a\int d^{10}x \left(V(f_{1})f'_{1}\sqrt{-\textrm{det}(\tilde{G}^{1}_{\alpha\beta})}+V(f_{2})f'_{2}\sqrt{-\textrm{det}(\tilde{G}^{2}_{\alpha\beta})}\right) \, .
\end{equation}
After performing the usual change of variables and using the descent relation between $T_9$, $T_8$ and $V$,  
we recognize this as being the $U(1) \otimes U(1)$ symmetric abelian DBI action for two separate D8-branes.

\section{Conclusions}
In this paper we have considered the generalization of Sen's tachyon condensation mechanism to the formation of two coincident BPS D8-branes on the world volume of tachyon kink-like configurations of two coincident non-BPS D9-branes. We found a natural extension of Sen's regularization of the singular tachyon kink profile, to the case of $U(2)$ tachyon valued field in the latter theory. What is apparent is the very different properties of the $Str$ vs $Tr$ prescription in taking the gauge trace in the non-abelian, non-BPS DBI action. The former leads to a series of very complicated terms that mix $D_\mu T, F_{\mu \nu}$ and more problematically individual $T$ in the tachyon potential $V(T)$. In particular, the latter consequence of taking $Str$ over gauge indices makes it very difficult to see tachyon condensation occurring in a way that is calculable and which yields the $Str$ prescription of the action of two coincident BPS D8-branes. 

Starting with the $Tr$ prescription however, we have explicitly shown that tachyon condensation gives rise directly to the
BPS action of two coincident D8-branes. This stark contrast between the $Str$ and $Tr$ prescriptions, parallels similar issues found by Garousi in \cite{Garousi:2007fn} regarding the existence (or not) of vortex solutions in brane-antibrane actions derived from coincident non-BPS D9-brane actions with $Tr$ or $Str$ prescriptions. 

Regarding further work in this area, firstly, it would be interesting to investigate non-abelian tachyon condensation, along the lines presented in this paper,
where one starts with e.g. two coincident non-BPS D$p$-branes with $p<9$. Then one expects to find the action of two coincident D$(p-1)$ BPS branes after tachyon condensation. The resulting action should presumably have the same structure as the one proposed by Myers \cite{Myers:1999ps}. Since the latter action is obtained via T-duality of the coincident D9-brane action, understanding the details of how non-abelian tachyon condensation works in this case would allow us to see if T-duality `commutes' with it.
On the other hand, since the Myers action has a $Str$ prescription, it is by no means obvious how one may realize such actions through the process of non-abelian tachyon condensation. 
Secondly, there are obvious extensions of our results to the case of multiple coincident non-BPS D9-branes and tachyon condensation leading to the action of multiple coincident BPS D8-branes. 
Finally, it would be interesting to show how one can inherit the correct Wess-Zumino terms for the BPS D$(p-1)$ branes from those that are part of the non-BPS action recently proposed in \cite{Garousi:2007fk,Garousi:2008ge}.
We hope to report further on these questions in the future.

$\phantom{123}$\\
{\bf Acknowledgments}
We would like to thank J. Ward for helpful discussions.
The work of V.C. is supported by a Queen Mary Westfield Trust Scholarship and G.T.by an EPSRC studentship.

%\bibliographystyle{JHEP-2}
%\bibliography{kink}

\providecommand{\href}[2]{#2}\begingroup\raggedright\endgroup

\end{document}